\documentclass[showpacs,preprintnumbers, pre]{revtex4}
\usepackage[dvips]{graphicx}
\usepackage{color}
\usepackage{epsfig}
\usepackage{dcolumn}
\usepackage{bm}
\usepackage{amssymb}
\usepackage{amsmath}

\usepackage[latin1]{inputenc}

\usepackage{soul}
\setstcolor{blue}

\newcommand{\be}{\begin{equation}}
\newcommand{\ee}{\end{equation}}
\newcommand{\ba}{\begin{eqnarray}}
\newcommand{\ea}{\end{eqnarray}}
\renewcommand{\vec}[1]{\mathbf{#1}}

\begin{document}

\title{Optimal search strategies of space-time coupled random walkers with finite lifetimes}
\author{D. Campos$^{1}$, E. Abad$^{2}$, V. M\'{e}ndez$^{1}$, S. B. Yuste$^{3}$ and K. Lindenberg$^{4}$}
\affiliation{Grup de F\'{\i}sica Estad\'{\i}stica, Departament de F\'{\i}sica, Facultat
de Ci\`{e}ncies, Universitat Aut\`{o}noma de Barcelona, 08193 Bellaterra
(Barcelona) Spain\\
$^{2}$ Departamento de F\'{\i}sica Aplicada and Instituto de Computaci\'on Cient\'{\i}fica Avanzada (ICCAEX) \\ Centro Universitario de M\'erida \\ Universidad de Extremadura, E-06800 M\'erida, Spain \\
$^{3}$Departamento de F\'{\i}sica and Instituto de Computaci\'on Cient\'{\i}fica Avanzada (ICCAEX), Universidad de Extremadura, E-06071 Badajoz, Spain \\
$^{4}$Department of Chemistry and Biochemistry, and BioCircuits Institute, University of California San Diego, 9500 Gilman Drive, La Jolla, CA 92093-0340, USA}

\begin{abstract}
We present a simple paradigm for detection of an immobile target by a space-time coupled random walker with a finite lifetime. The motion of the walker is characterized by linear displacements at a fixed speed and exponentially distributed duration, interrupted by random changes in the direction of motion and resumption of motion in the new direction with the same speed. We call these walkers "mortal creepers". A mortal creeper may die at any time during its motion according to an exponential decay law characterized by a finite mean death rate $\omega_m$. While still alive, the creeper has a finite mean frequency $\omega$ of change of the direction of motion. In particular, we consider the efficiency of the target search process, characterized by the probability that the creeper will eventually detect the target.   Analytic results confirmed by numerical results show that there is an $\omega_m$-dependent optimal frequency  $\omega=\omega_{opt}$ that maximizes the probability of eventual target detection. We work primarily in one-dimensional ($d=1$) domains and examine the role of initial conditions and of finite domain sizes. Numerical results in $d=2$ domains confirm the existence of an optimal frequency of change of direction, thereby suggesting that the observed effects are robust to changes in dimensionality. In the $d=1$ case, explicit expressions for the probability of target detection in the long time limit are given. In the case of an infinite domain, we compute the detection probability for arbitrary times and study its early- and late-time behavior. We further consider the survival probability of the target in the presence of many independent creepers beginning their motion at the same location and at the same time. We also consider a version of the standard ``target problem" in which many creepers start at random locations at the same time.
\end{abstract}


\maketitle

\section{Introduction}

There are many processes in nature in which one or more \textit{randomly moving entities} detect one or more \textit{targets}.  Among many examples we mention binding or trapping processes in molecular environments, diffusion-limited chemical reactions, and the search for nutrients in a biological context such as predator-prey situations in ecological systems. In many of these ``target problems" the question of interest is the efficiency with which the mobile entity finds the target as measured by the time it takes to accomplish this outcome. The mathematical quantities to be calculated for this purpose are first passage time distributions or their moments, especially the mean first passage time (MFPT) for an entity to reach the target. The above problem is then tantamount to the calculation of the survival probability of the target. Many such problems for moving entities that execute random walks or, essentially equivalently, diffusive motion, have been presented in ref. \cite{Redner}. Recently, the role of parallel and intermittent search strategies in the above context has also attracted considerable interest (see e.g. refs. \cite{EliEtAl2, EliEtAl3}).

Other types of motion have frequently been considered in the context of target capture, especially in the ecology literature \cite{MCBBook}, including L\'evy flights and L\'evy walks (see e.g. chapters 7 and 8 in ref. \cite{SokoKlafter2011} and a recent review on L\'evy walks \cite{ZabDenKl2015}). L\'evy flights are characterized by instantaneous steps whose lengths are chosen from a probability distribution that decays as a power law.  This leads to scale-free superdiffusive motion. L\'evy walks differ from L\'evy flights in that the steps are not instantaneous but occur at a constant speed. Other models of superdiffusion are based on step-step correlations \cite{EliEtAl1} and have also been considered in the target capture problem (see e.g. the example with fractional Brownian motion cited in \cite{BMS2013}). Whereas normal diffusion is associated with a mean square displacement that grows linearly with time, $\langle x^2\rangle\sim t$, in superdiffusive motion the growth is superlinear. At the other extreme lie subdiffusive models, ones in which the growth is sublinear. These are often associated with crowded environments where motion is impeded by traps, barriers, and dead ends. Recent work has extended target capture models to subdiffusive motions arising from continuous time random walks (CTRWs) \cite{YL2007, OurBookChapter2011, BAY2009, FrankeMajumdar}.

Stepping back from the outcomes of various groups of models in terms of the mean square displacement to the mesoscopic basis for these outcomes, we can broadly characterize the random motions by two distributions:  the distribution of stepping times (in the language of continuous time) and the distribution of stepping lengths (in the language of continuous space). Normal diffusion appears if the first moment of the  waiting  time distribution and the two first moments of the stepping length distribution are finite. The ubiquitous example is an exponential distribution of stepping times and a Gaussian distribution of stepping lengths. Subdiffusive motion can be achieved with stepping time distributions that lack a first moment, that is, there is a strong probability of a long sojourn at one location  before moving on, so much so that the mean stepping time diverges. Superdiffusive motion, on the other hand, is associated with distributions of stepping distances that lack a second moment, that is, the variance of the step length diverges. We note that in recent years work on target capture has been extended to more complex processes such as, for instance, ones displaying non-trivial spatial or space-time correlations. One such example is given by the capture of an evasive prey which is able to detect predators within a certain range and step back accordingly \cite{EvasivePrey}.

A feature of our own recent work is the recognition that in many problems the moving entity has a finite lifetime \cite{AYL2010, SurvivalEvanescent1d, SurvivalEvanescent, ourPRL, evCTRW, AYLJMMNP, MortalBookChapter, NovaBC}. In general, this finite lifetime is randomly distributed.  We have considered a variety of lifetime distributions, from exponential to algebraic.  The former is the distribution associated with a ``unimolecular"concentration decay process $entity \to 0$, while the latter may be indicative of more complex, encounter-controlled kinetics leading to death or inactivation \cite{PrivmanBook}.

Returning to the target problem, the goal of a great deal of work, whichever the underlying model of motion and death, is to establish if optimization is possible. The  goal may be to maximize the probability of encountering the target (lowest survival probability) or to minimize it (greatest survival probability).  The latter case is for instance relevant in ecology, e.g., if a moving prey is trying to avoid a predator. Apart from the aforementioned details of the stepping time and the stepping length distributions, the relevant parameters with respect to which the search process must be optimized vary with the details of the model, for example initial conditions, boundary conditions, size, or dimensionality of the system. In particular, the literature refers to the \emph{target problem} when there are many walkers starting their walk at random locations.  There is a vast literature on the target problem \cite{Redner, BMS2013}, but not on the target problem with mortal rather than immortal walkers. However, in recent days the literature on the latter topic and related problems is increasing rapidly \cite{SurvivalEvanescent1d, SurvivalEvanescent, ourPRL, evCTRW, AYLJMMNP, MortalBookChapter, NovaBC, BeniRedner, MeersonRedner}.

In any case,  interesting effects arise from the interplay of the lifetime of an entity and the time it takes it to reach the target.  If the  lifetime is much shorter than the time it would take an immortal entity to reach the target, then the problem is in a sense uninteresting because the target is simply hardly ever  reached.  In the reverse case, the fact that the entity has a finite lifetime does not matter because it most frequently reaches the target before it dies.  Clearly, the most interesting scenarios arise when these times are comparable.  Since these times are both distributed, the most interesting cases are those where there is a significant overlap of the two distributions.

In this paper we examine the search efficiency of mortal L\'evy walkers whose step duration and step length are exponentially distributed.  More explicitly, the motion of our walkers is characterized by ballistic displacements at a \textit{fixed speed}, interrupted at random times by random changes in the direction of motion and resumption of motion in the new direction with the same speed as before. Following Hughes \cite{Hughes}, we call these walkers ``creepers'' to distinguish them from ``leapers'', that is, walkers performing instantaneous jumps. In addition, we assume that our creepers are mortal, i.e., they have a finite lifetime which we assume to be exponentially distributed here.

In investigating the survival probability of the target and the optimization (in this work, minimization) thereof, the interplay of two parameters is of special interest, namely, the average death rate $\omega_m$ and the average frequency of reorientation $\omega$ of the creeper.  We also emphasize an important point of the model: the creeper may reach the target at any time during its trajectory, which need not coincide with the end of a linear displacement. Most (but not all) of our paper deals with this problem in one dimension, where reorientation is simply a direction reversal.  This case is admittedly considerably simpler than the problem in higher dimensions, but we do show numerically that the results are qualitatively similar in two dimensions.

We organize our presentation as follows.  In Sec.~\ref{sec2} we set forth the formalism for the motion of a single mortal creeper as well as for a collection of mortal creepers.  In Sec.~\ref{sec3} we use this formalism to calculate the survival probability of a target in the one-dimensional case, both in an infinite and in a finite domain. In the former case we give an explicit expression valid for arbitrary times. Section~\ref{sec4} presents a brief numerical study of the two-dimensional problem.  Finally, in Sec.~\ref{sec5} we summarize our main findings and possible extensions of this work.

\section{General framework}
\label{sec2}

Consider a system consisting of an immobile, impenetrable hyperspherical target of radius $R$ and a randomly moving creeper of negligible spatial extent initially located at a certain distance from the target surface. The target is detected by the searcher as soon as the latter intersects the target surface. Our goal is to calculate the survival probability of the target.

\subsection{Master equation describing the motion of the creeper}

The motion of the creeper in one dimension starting from initial position $x_0$ is described by the equation
\begin{equation}
j(x,t|x_{0})=\int_{0}^{t}dt^{\prime }\int_{-\infty }^{\infty }dx^{\prime
}j(x-x^{\prime },t-t^{\prime }|x_{0})\Psi (x^{\prime },t^{\prime
})+j_{0}.  \label{1}
\end{equation}
Here $j(x,t|x_{0})dx\,dt$ is the probability that the creeper completes a displacement and thus
performs a turn (direction reversal) during the infinitesimal time interval $[t,t+dt]$ while dwelling in the infinitesimal segment $[x,x+dx]$, and
$\Psi (x,t)$ is the probability density of having performed a displacement $x$ in a time interval $t$ between consecutive turns. Thus, Eq. (\ref{1}) represents a detailed bookkeeping of the possible ways in which a creeper may finish a displacement at a given time
in terms of all the possible positions and times of the previous turn. The deterministic initial condition is embodied in the choice $j_{0}\mathbf{(}x)=\delta(x-x_0)\delta(t)$. In quantities that are conditional on the initial condition the dependence on initial time $t=0$ is understood, e.g., $j(x,t|x_0)$ is an abbreviated notation for $j(x,t|x_0,0)$.

The description of the motion given by (\ref{1}) is complemented by the equation
\begin{equation}
p(x,t|x_{0}) =\int_{0}^{t}dt^{\prime }\int_{-\infty }^{\infty }dx^{\prime
}j(x-x^{\prime },t-t^{\prime }|x_{0})\phi (x^{\prime },t^{\prime }),
\label{pjeq}
\end{equation}
which relates the statistical properties of the location of the creeper at time $t$ to those of its sojourn in different regions of space for different periods of time. The quantity $p(x,t|x_{0})\,dx$ is the probability of finding the walker inside the interval $[x,x+dx]$ at time $t$ provided it was at $x_0$ at time $t=0$, and $\phi (x,t)\,dx$ is the probability that a step remains unfinished after a time interval $t$ since its start, during which time a displacement $x$ was performed (if the creeper velocity is $v_f>0$, then obviously $\phi \propto \delta(x\pm v_ft)$ - see below).

The particular case of a walk with an exponentially decaying waiting time pdf is described by the following joint density:
\begin{equation}
\Psi (x,t)=\frac{1}{2}\left[ \delta (x+v_{f}t)+\delta (x-v_{f}t)\right]
\omega\, e^{-\omega t},
\label{pdf1d}
\end{equation}
where, as introduced earlier,  $\omega$ is the  average rate at which displacements are terminated by a turn, implying that $\omega^{-1}$ is the mean duration of a displacement in one direction.  We also call this the characteristic ``persistence time'' of the creeper. The product form of $\Psi(x,t)$ reflects the fact that the speed of each displacement is independent of its duration. For simplicity, we assume that the beginning of the first displacement coincides with the initial time $t=0$, and that the creeper initially has equal probability to move to the left or to the right. The $\delta$-functions in fact indicate the constant modulus $v_f$ of the creeper's velocity. Given the density in Eq. \eqref{pdf1d}, the corresponding expression for the density $\phi (x,t)$ is obtained by computing the probability $\int_t^\infty \omega\, e^{-\omega t'} \, dt'$ that a given step is unfinished after a time $t$:
\begin{equation}
\phi(x,t)=\frac{1}{2}\left[ \delta (x+v_{f}t)+\delta (x-v_{f}t)\right]
\int_t^\infty \omega\, e^{-\omega t'} \, dt' =\frac{1}{2}\left[
\delta (x+v_{f}t)+\delta (x-v_{f}t)\right]  e^{-\omega t } = \frac{ \Psi(x,t) }{ \omega }.
\label{surv}
\end{equation}
Equations~(\ref{1}-\ref{surv}) provide a full description of the motion of an immortal creeper.

We now introduce a finite lifetime for the creeper.
We take the death process to be independent of the transport properties.
The sojourn pdf $p^{*}(x,t|x_{0})$ for a mortal creeper can then be expressed in terms of the pdf $p(x,t|x_{0})$ for an immortal creeper as
\begin{equation}
p^{*}(x,t|x_{0})=\varphi(t) p(x,t|x_{0}),
\label{mortal}
\end{equation}
where the mortality function $\varphi(t)$ is the probability that the creeper has survived for a time interval $t$ since it started moving, and $\varphi(0)\equiv 1$. Throughout the remainder of this work we will use asterisks to denote quantities describing mortal creepers.

\subsection{Probability of target detection}

We next turn to the calculation of the survival probability of a target centered at $x_{tg}$, that is, the probability that it has not been detected by a creeper. In the one-dimensional case we may assume without loss of meaningful information that the target radius is vanishingly small, $R=0$.  We first write the relevant equations for the case of immortal creepers and then show how to extend them to the case of mortal creepers.

\subsubsection{Single immortal creeper}

We introduce $q(x_{tg},t|x^\prime, t^\prime)$ as the probability per unit time for the creeper to detect the target (that is, the target detection rate) at time $t$ no matter how often this detection has occurred before, given that the creeper is found at $x^\prime$ at time $t^\prime <t$. In particular, $q(x_{tg},t|x_{0},0)\equiv q(x_{tg},t|x_{0})$ is the detection rate given that the creeper started its search at $x_{0}$ at the initial time $t=0$. To detect the target, the creeper must step on location $x_{tg}$. Therefore $q(x_{tg},t|x_{0})$ is clearly proportional to the probability density $p(x_{tg},t|x_{0})$ introduced in the previous section (the special case with $x_0=x_{tg}$ will be dealt with separately). In what follows we assume perfect target detection, that is, the creeper detects the target with unit probability whenever it steps on location $x_{tg}$.

We are ultimately only interested in the first encounter of the creeper with the target. The rate $f(x_{tg},t|x_{0})$ of \emph{first} detection (probability per unit time to detect the target \emph{for the first time})
for a creeper that starts at $x_0$ can be computed as a function of the overall detection rate $q(x_{tg},t|x_{tg},t^\prime)$ from the renewal equation~\cite{campos13}:
\begin{equation}
q(x_{tg},t|x_{0})=f(x_{tg},t|x_{0})+\int_{0}^{t} q(x_{tg},t|x_{tg},t^{\prime}) f(x_{tg},t^{\prime}|x_{0}) \, dt^{\prime }.  \label{2a}
\end{equation}
Here the rate at which the target is detected at time $t$  has been decomposed into two mutually exclusive contributions. One contribution arises from first detection events at time $t$, given by the first term on the right hand side.  The second contribution on the right hand side accounts for revisitations to the target location, that is, first visitations of the target at time $t'<t$ followed by a return to the target at time $t$ (following any number of additional revisitations in between). Hence the first detection rate also appears in this term.

Since the system is invariant with respect to a time shift, one has that $p(x_{tg},t|x^{\prime},t^{\prime})=p(x_{tg},t-t^{\prime}|x^\prime)$, and thus $q(x_{tg},t|x^{\prime},t^{\prime})= q(x_{tg},t-t^{\prime}|x^{\prime})$. One therefore obtains the closed equation
\begin{equation}
q(x_{tg},t|x_{0})=f(x_{tg},t|x_{0})+\int_{0}^{t} q(x_{tg},t-t^{\prime}|x_{tg}) f(x_{tg},t^{\prime}|x_{0}) \, dt^{\prime }.  \label{conveq}
\end{equation}
This is especially convenient because the integral is now a convolution so that the equation is greatly simplified in Laplace space.
The Laplace transform of Eq.~(\ref{conveq}) readily leads to
\begin{equation}
f(x_{tg},s|x_{0})=\frac{q(x_{tg},s|x_{0})}{1+q(x_{tg},s|x_{tg})}.
\label{relfq}
\end{equation}
Note that we use the same notation for any function of time and its Laplace transform.  The choice is made clear by the argument. The mean first detection time of the target (hereafter called the mean first passage time, MFPT) can then be obtained from this result for the Laplace transform of the probability per unit time of first detection, cf. \cite{campos13}.

We focus instead on the closely related survival probability of the target, which we assume to be killed as soon as the creeper arrives at the target location for the first time. The  survival probability of the target is then given by
\begin{equation}
S(t)= 1-\int_{0}^{t }f(x_{tg}, t^\prime |x_{0}) dt^\prime.
\end{equation}
In Laplace space, one then has 
\begin{equation}
S(s)= \frac{1}{s}-\frac{1}{s}\,f(x_{tg}, s |x_{0}).
\label{laptransS}
\end{equation}
In particular, the asymptotic survival probability $S_\infty \equiv S(t\to\infty)$ is then given by
\begin{equation}
S_{\infty }= 1-\int_{0}^{\infty }f(x_{tg},t|x_{0}) dt =1-\lim_{s\rightarrow 0} f(x_{tg},s|x_{0}).
\label{targetdies}
\end{equation}
Since the searcher is immortal and the embedding geometry is one-dimensional, the motion is recurrent, that is, every location is guaranteed to be visited. This ensures that the target will eventually be detected with certainty, i.e.,  the integral on the right hand side of Eq.~(\ref{targetdies}) is unity.
These arguments can easily be generalized to the case of an arbitrary initial probability distribution $p(x_0)$ by taking the corresponding average in
Eq.~\eqref{targetdies}.

\subsubsection{Single mortal creeper}

Equation (\ref{2a}) must be modified  for mortal creepers since
detection is now conditional on the survival function $\varphi(t)$ of the creeper.
To take this into account, we multiply Eq.~(\ref{2a}) by $\varphi(t)$ and rewrite the resulting equation
as follows:
\begin{equation}
q^{\ast}(x_{tg},t|x_{0})=f^{\ast}(x_{tg},t|x_{0})+\int_{0}^{t}q^{\ast}(x_{tg},t|x_{tg},t^{\prime })f^{\ast}(x_{tg}, t^{\prime}|x_{0})\, dt^{\prime }.
\label{mortalbalance}
\end{equation}
Here
$q^{\ast}(x_{tg},t|x_{tg}, t^\prime)\equiv [\varphi(t)/\varphi(t')]\,q(x_{tg},t|x_{tg}, t^\prime)$
and
 $f^{\ast}(x_{tg},t|x_{0})\equiv [\varphi(t)/\varphi(0)]\,f(x_{tg},t|x_{0})=\varphi(t)\,f(x_{tg},t|x_{0})$.
The quantities $q^\ast$ and $f^\ast$ have the same interpretation as $q$ and $f$, but now weighted with the appropriate conditional survival
probabilities. In particular, the ratio $\varphi(t)/\varphi(t^\prime)$ is the survival probability of the creeper in the time interval $[t^\prime, t]$ conditional on having survived up to time $t^\prime$.

Since $\varphi(t)/\varphi(t^\prime)$ depends on both  $t$ and $t^\prime$ separately and in general not on the time difference
$t-t^\prime$, the death process in general destroys the invariance of the propagator with respect to a time shift.
It is then not possible to obtain a closed equation of the form (\ref{conveq}) for the quantities
with an asterisk. However, if the survival function decays exponentially, $\varphi(t)=e^{-\omega_{m} t}$, one has $\varphi(t)/\varphi(t^\prime)=e^{-\omega_{m} (t-t^\prime)}$ and the shift invariance is preserved~\cite{ourPRL}.
For this particular case, one does find a renewal equation similar to (\ref{conveq}), namely,
\begin{equation}
q^\ast(x_{tg},t|x_{0})=f^\ast (x_{tg},t|x_{0})+\int_{0}^{t} q^\ast (x_{tg},t-t^{\prime}|x_{tg}) f^\ast (x_{tg},t^{\prime}|x_{0}) \, dt^{\prime }.
\end{equation}
In analogy to Eq.~(\ref{relfq}), we find in Laplace space $f^{\ast }(x,s|x_{0})=q^\ast (x_{tg},s|x_0)[1+q^\ast (x_{tg},s|x_{tg})]^{-1}$, and applying the shift theorem for the Laplace transform we obtain
\begin{equation}
\label{fmort}
f^{\ast }(x_{tg},s|x_{0})=\frac{q(x_{tg},s+\omega _{m}|x_{0})}{
1+q(x_{tg},s+\omega _{m}|x_{tg})}=f(x_{tg},s+\omega _{m}|x_{0}).
\end{equation}
The survival probability at time $t$ in the presence of an exponential mortality function then is
\begin{equation}
S^{\ast}(t)=  1-\int_{0}^{t }f^{\ast}(x_{tg},t^\prime |x_{0})\, dt^\prime,
\end{equation}
whose Laplace transform yields
\begin{equation}
S^{\ast}(s)=  \frac{1- f^{\ast}(x_{tg},s|x_{0})}{s}.
\end{equation}
The asymptotic value of the survival probability is formally similar to that found in the absence of mortality,
\begin{equation}
S_{\infty }^\ast  =1-\int_{0}^{\infty }f^\ast(x_{tg},t|x_{0}) dt =1-\lim_{s\rightarrow 0}f^\ast (x_{tg},s|x_{0}),
\end{equation}
but it no longer vanishes.  In particular, we now obtain the final result
\begin{equation}
S_{\infty }^{\ast}=
1-\frac{q(x_{tg},s=w_m|x_{0})}{1+q(x_{tg},s=w_m|x_{tg})}.
\label{mortalSP}
\end{equation}
Once again these arguments can easily be generalized to an arbitrary initial distribution $p(x_0)$.

\subsubsection{$N$ immortal or mortal creepers}

Finally, suppose that instead of a single creeper we now have a collection of $N$ statistically independent immortal creepers all starting at the same location $x_0$ at time $t=0$.  The survival probability $S_N(t)$ of the target up to time $t$ is simply the probability that it has not been detected by any of the searchers up to that time. We now have $S_N(t)=[S(t)]^N$, and also the asymptotic result $S_{N,\infty}=S_\infty^N$ . The same reasoning applies to mortal creepers provided they all have the same mortality function, i.e.,  $S_{N,\infty}^{\ast}=(S_\infty^{\ast})^N$.

\section{Explicit results in one dimension}
\label{sec3}
In this section we obtain a variety of explicit results for the survival probability of a target in the presence of one or more mortal and immortal creepers in dimension $d=1$, and compare a number of them with the results of numerical simulations.

\subsection{Infinite domain}

\subsubsection{Single immortal creeper}

We begin with a single creeper and a target at location $x_{tg}$. The convolution structure of Eqs.~(\ref{1}) and (\ref{pjeq}) translates into a Fourier-Laplace transform  $p(k,s|x_{0})\equiv {\cal F}\left\{p(x,t|x_0)\right\}$ of the propagator which satisfies the so-called Montroll-Weiss equation (see e.g. \cite{SokoKlafter2011}, p. 112):
\begin{equation}
p(k,s|x_{0})=\frac{\phi(k,s) e^{-ikx_{0}} }{1-\Psi(k,s)}.
\label{montrollweiss}
\end{equation}
We adopt the convention $f(k)=\int_{-\infty}^\infty dx f(x)e^{ikx}$ for the Fourier transform, using the same letter $f(k)$ for the Fourier transform as we do for the space-dependent function $f(x)$, again relying on the argument to make the distinction.
In the particular case of our Markovian walk the above equation takes the form  [see Eq. (\ref{surv})]
\begin{equation}
p(k,s|x_{0})=\frac{\omega^{-1} e^{-ikx_{0}} }{ \Psi^{-1}(k,s)-1 }.
\label{markov}
\end{equation}

Next, taking the Fourier-Laplace transform of the joint distribution $\Psi(x,t)$ defined by Eq. (\ref{pdf1d}), one obtains
\begin{equation}
\Psi(k,s)=\frac{w}{2}\,\left[\frac{1}{s+w+ikv_f}+\frac{1}{s+w-ikv_f}\right]=\frac{(s+w)w}{(s+w)^2+v_f^2 k^2}.
\end{equation}
Inserting this expression into Eq. (\ref{markov}) yields
\begin{equation}
p(k,s|x_{0})=\frac{(w+s) e^{-ikx_0}}{s(s+w)+v_f^2 k^2} .
\end{equation}
As is well known, the above expression is simply the Fourier-Laplace transform of the free solution of the telegrapher's equation (see e.g. \cite{BookWeiss}, p. 246):
\begin{equation}
v_f^2 \frac{\partial^2}{\partial x^2}p(x,t|x_0)=\frac{\partial^2}{\partial t^2}p(x,t|x_0)+w\,\frac{\partial}{\partial t}p(x,t|x_0)
\end{equation}
(the telegrapher's equation can also be
obtained directly  by differentiating the set of equations (\ref{1})-(\ref{pjeq}) with respect to time).
Fourier inversion of $p(k,s|x_0)$ yields
\begin{equation}
p(x,s|x_0)=\frac{1}{2v_f}\,\sqrt{\frac{s+w}{s}}\,e^{ -\sqrt{s(s+w)}|x-x_0|/v_f}.
\label{LTp}
\end{equation}

Finally, introducing the definition $\alpha_w(s)\equiv \sqrt{s(s+w)}$ one has
\begin{equation}
p(x,s|x_{0})=\frac{1}{2v_f}\,\frac{\alpha_w(s)}{s}\,e^{ -\alpha_w(s)|x-x_{0}|/v_f}.
\label{LTp2}
\end{equation}
The normalization of this probability density to unity can easily be ascertained through $\int_{-\infty}^\infty p(x,s|x_{0})\,dx=s^{-1}$. The analytic inversion with respect to the Laplace variable $s$ can also be carried out and leads to an expression in terms of modified Bessel functions. However, bearing in mind that we are focusing on the survival probability, it is more convenient to continue working in Laplace space.

As an aside, we note that the expression (\ref{LTp}) for the Laplace-transformed propagator can be compared with the one for normal diffusion, $p_{nd}(x,s|x_0)$. One has
\begin{equation}
p_{nd}(x,s|x_0)\equiv {\cal L}\left\{\frac{e^{-(x-x_{0})^2/(4Dt)}}{{\sqrt{4\pi Dt}}}\right\}=\frac{e^{-\sqrt{s/D}|x-x_{0}|}}{2\sqrt{Ds}},
\end{equation}
which coincides with the small $s$ limit of Eq.~(\ref{LTp}) if one takes $D=v_f^2 w^{-1}$. This in turn implies a coincidence at long times.

The rate at which an immortal creeper detects the target at $x_{tg}$ any number of times is the sum of the probability fluxes of particles that arrive at $x_{tg}$ from the left (with velocity $v_f$), $\Phi_+(x_{tg},t)$,  and from the right (with velocity $-v_f$), $\Phi_-(x_{tg},t)$. In terms of differential detection probabilities one has
\begin{equation}
q(x_{tg},t \vert x_{0})\, dt= \Phi_+(x_{tg},t)dt +\Phi_-(x_{tg},t)dt =
\int_{x_{tg}-v_{f}dt}^{x_{tg}} p(x,t;v_{f} \vert x_{0}) dx + \int_{x_{tg}}^{x_{tg}+v_{f}dt} p(x,t;-v_{f} \vert x_{0}) dx,
\label{qleftright2}
\end{equation}
where $p(x,t;v_{f} \vert x_{0})dx$ and $p(x,t;-v_{f} \vert x_{0})dx$ are, respectively, the ensemble probabilities that creepers at a position $x$ at time $t$ move toward the right and toward the left. For $t> 0$ an expansion in $dt$ yields
\begin{equation}
q(x_{tg},t \vert x_{0}) = v_{f}  \left[ p(x_{tg},t, v_{f} \vert x_{0}) + p(x_{tg},t, -v_{f} \vert x_{0}) \right] = v_{f} p(x_{tg},t \vert x_{0}).
\label{qleftright3}
\end{equation}
Obviously, $q(x_{tg},0 \vert x_{0})=0$ when $x_0\neq x_{tg}$. However, the case of $t=0$ and $x_0=x_{tg}$ must be treated with care. First, we realize that $p(x,0;\pm v_f|x_0)=(1/2) \delta(x-x_0)$ so that, for $t\to 0$, half of the creepers go to the right from $x_0$,  whereas none go to $x_0$ from the right because  there are no creepers beyond $x_0$. In other words, for $t\to 0$ position $x_0$ is special: all the creepers move away from $x_0$, and no creeper goes toward $x_0$. This means that
\begin{equation}
q(x_{0},0 \vert x_{0})\,dt=  \Phi_{\pm}(x_0,0)dt =
\frac{1}{2}.
\end{equation}
However, $ v_{f} p(x_{0},0 \vert x_{0})dt
=\Phi_{+}(x_0,0)dt+\Phi_{-}(x_0,0)dt
=1$, so that we can rewrite $q(x_{0},0 \vert x_{0})\,dt=v_{f} p(x_{0},0 \vert x_{0})dt-1/2$. On the other hand, $p(x_{tg},0 \vert x_{0})=0$ for $x_0\neq x_{tg}$. Therefore, these results together with Eq.~\eqref{qleftright3} can be written as follows \cite{campos14b}:
\begin{equation}
q(x_{tg},t|x_0) = \begin{cases}
v_{f}p \left( x_{tg} ,t \mid x_{0} \right), & x_0 \neq x_{tg} \\[2mm]
v_{f}p \left( x_{tg}, t \mid x_{0} \right) -\frac{1}{2}\delta (t), & x_0=x_{tg}
\end{cases}
\label{qrate}
\end{equation}
Note that we have assumed that the walker does not detect the target at $t=0$ even if its initial location is $x_0=x_{tg}$. Detection only occurs if the creeper returns to this location.  This is not a necessary assumption, and one could easily deal with inclusion of immediate target death if the initial creeper location is at the target.

We close this part by noting that taking the Laplace transform of Eq. \eqref{qrate} yields a straightforward relation between $q(x_{tg},s|x_0)$ and the unrestricted Laplace-transformed propagator $p(x,s|x_{0})$ of the telegrapher's equation evaluated at $x=x_{tg}$. The resulting closed analytic expression can then be used to explicitly compute $f(x_{tg},s|x_0)$ by means of Eq. \eqref{relfq}. Inserting the resulting expression into Eq. \eqref{laptransS} yields an expression for the Laplace-transformed detection probability which can subsequently be inverted to explicitly compute $S(t)$. Formally, one may assume that the target is killed as soon as it is detected. The detection probability can then be identified with the survival probability of the target. If one assumes that the searcher also dies as soon as it detects the target, the target can be regarded as a trap and the survival probability of the searcher then also becomes identical to $S(t)$. The computation of the survival probability of an immortal walker whose motion is governed by the telegrapher's equation in the presence of a trap was carried out in ref. \cite{MasPorrWeiss}, and the result is expressible in terms of modified Bessel functions (also see the remarks at the end of \ref{TDep} ).

\subsubsection{Single mortal creeper}

We next calculate $S_\infty^*$,  the asymptotic survival probability of the target in the presence of a mortal creeper.
To do this we refer back to Eq.~(\ref{mortalSP}). The right hand side requires the Laplace transform of Eq.~(\ref{qrate}), which is easily calculated using Eq.~(\ref{LTp2}). We readily obtain
\begin{equation}
S_{\infty }^{\ast }= \begin{cases}
1-\dfrac{ \alpha_\omega (\omega_m)  }{\alpha_\omega (\omega_m)+\omega _{m}}
\;e^{-\alpha_\omega(\omega_m)\left\vert x_{tg}-x_{0}\right\vert
/v_{f}},
& x_0 \neq x_{tg} \\[4mm]
\dfrac{2\omega _{m}}{\alpha_\omega (\omega_m)+\omega _{m}}, &  x_{0}=x_{tg}
\end{cases}
\label{survivalinf}
\end{equation}
This can be compared to the corresponding normal diffusion model in the limit $v_{f} \rightarrow \infty$, $\omega \rightarrow \infty$, with fixed $D \equiv v_{f}^{2} \omega^{-1}$. In this case one obtains the expression
\begin{equation}
S_{nd,\infty }^{\ast }=
 1-e^{-\sqrt{\omega _{m} /D } \left\vert x_{tg}-x_0 \right\vert}
\end{equation}
for all $x_0$.
Note that in the limit of infinite lifetime, $\omega_m\to 0$, all of these survival probabilities lead (as they should) to the result $S^*_\infty=S^*_{nd,\infty}\to S_\infty=0$, that is, the target does not survive regardless of the initial position of the creeper.  On the other hand, in the limit of a creeper that dies infinitely quickly, the  target survives with unit probability, $S^*_\infty=S^*_{nd,\infty}=1$, because the creeper essentially does not move before dying.  This is the case even when $x_0=x_{tg}$ since our convention is not to count the initial position as a step in the process.

\begin{figure}
\includegraphics{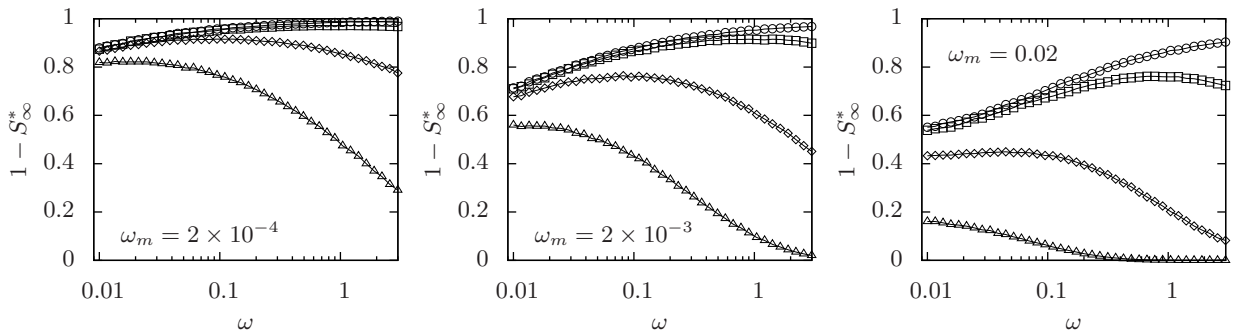}
\caption{Plot of the search efficiency in an infinite domain as a function of the mean rate $\omega$ of direction reversal of the creeper for different initial conditions $x_0$ and different values of the mortality rate $\omega_m$: $\vert x_{tg}-x_{0} \vert=0.1$ (circles), $ 1$ (squares), $10$ (diamonds) and $50$ (triangles). The solid lines correspond to the analytic result (\ref{effinf}), whereas the points denoted by symbols are obtained numerically from the average over the trajectories of $10^{5}$ independent random searchers. In all cases we use $v_{f}=1$. The domain size in the simulations has been chosen large enough for finite-size effects to be negligible.}
\label{figura1}
\end{figure}

Comparisons of our results with numerical simulations are presented in terms of the search efficiency $1-S_\infty^*$, which we write explicitly because this makes the comparison more straightforward:
\begin{equation}
1-S_{\infty }^{\ast }=
\begin{cases}
\dfrac{\alpha_\omega (\omega_m) }{\alpha_\omega (\omega_m)+\omega _{m}} \; e^{-\alpha_\omega(\omega_m)\left\vert x_{tg}-x_{0}\right\vert
/v_{f}},
& \quad x_{0}\neq x_{tg} \\[4mm]
\dfrac{\alpha_\omega (\omega_m)-\omega _{m}}{\alpha_\omega (\omega_m)+\omega _{m}}, & \quad x_{0}=x_{tg}
\end{cases}
\label{effinf}
\end{equation}
Figure \ref{figura1} shows three plots of the search efficiency as a function of the mean frequency $\omega$ of change of direction of the creeper (direction reversal in dimension $d=1$) for different values of the creeper mortality rate $\omega_m$  and for different initial conditions. The analytic result (\ref{effinf}) is in excellent agreement with results from Monte Carlo simulations averaged over $10^{5}$ realizations of the random walk.

As would be expected, the search efficiency increases with decreasing values of the initial distance $|x_{tg}-x_0|$ and decreasing mortality rate (in the latter case, the searcher obviously has a longer time to detect the target). A number of the  plots in Figure \ref{figura1} show an optimal value of $\omega$ for which the search efficiency becomes maximal. This is the result of a trade-off between maintaining persistence long enough to avoid spending a great deal of time in a region devoid of the target, and breaking persistence often enough to avoid large departures from a nearby target. Obviously, the details on how this trade-off works depend on the characteristic time $\omega^{-1}$ between turns, on the mean lifetime $\omega_m^{-1}$ of the creeper, and on the typical time $\tau_0=|x_0-x_{tg}|/v_f$ it would take the creeper to cover the initial distance ballistically. In a finite domain of linear size $L$ an additional time scale is introduced, namely the typical time needed to cover the full system length in the absence of the target $L/v_f$.  This is thus a complex problem, but Figure 1 shows the fairly universal feature of an optimal search efficiency (minimum survival probability of the target).

\begin{figure}
\includegraphics{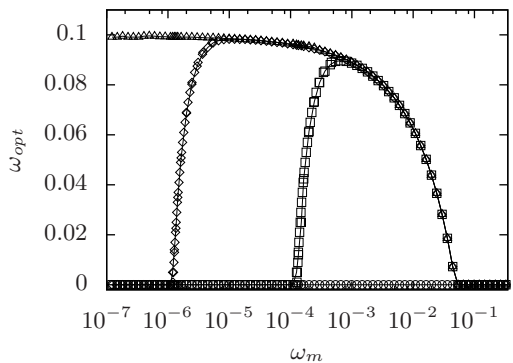}
\caption{Behavior of the optimal persistence $\omega_{\text{opt}}$ as a function of the mortality rate $\omega_{m}$  for different values of system size $L$, $L=100$ (circles), $10^{3}$ (squares), $10^{4}$ (diamonds) and $L \rightarrow \infty$ (triangles). Solid lines correspond to the analytic result (\ref{survivalfinite}), while symbols denote data points obtained from Monte Carlo simulations (averaging over the trajectories of $10^{5}$ independent mortal creepers). In all cases we place the target at $x_{tg}=0$, begin the walk of the creeper at $x_{0}=10$ and the creeper has speed $v_{f}=1$.}
\label{figura3}
\end{figure}

Figure~\ref{figura3} shows the optimal frequency of turns (optimal persistence) as a function of creeper mortality rate for various values of $L$. While we leave most of the discussion of a finite domain to subsection \ref{subs:findom}, it is helpful to consider the $L\to\infty$ behavior shown in Figure~\ref{figura3}. For very large mortality rates it is clear that the creeper must perform ballistic motion in order to have a chance to detect the target. However, as the mortality rate becomes smaller, it may be desirable to change direction a few times in order to avoid long excursions away from the target, thereby leading to a finite optimal frequency of turns $\omega_{opt}$. The relation between $\omega_{opt}$ and $\omega_m$ can then be expressed as follows:
\begin{equation}
\label{x}
\omega_m=
\frac{(1-\omega_\text{opt}|x_{tg}-x_0|/v_f)^2}
{|x_{tg}-x_0|/v_f\,(2-\omega_\text{opt}|x_{tg}-x_0|/v_f)}.
\end{equation}
From here we see that $\omega_\text{opt}=v_f/|x_{tg}-x_0|$ when $\omega_m\to 0$, and also that the threshold value of the mortality rate for which $\omega_\text{opt}$ becomes zero is $\omega_m=v_f/(2 |x_{tg}-x_0|)$, in agreement with Figure~\ref{figura3} and with the description given above. We return to Figure~\ref{figura3} later to discuss the case of finite $L$.

\subsubsection{Time dependence}
 \label{TDep}

Results for the search efficiency $1-S^{\ast}(t)$ at finite times can be obtained by inversion of the exact expression in the Laplace domain. The latter reads

\begin{equation}
\frac{1}{s}-S^{\ast}(s)=
\begin{cases}
\dfrac{1}{s}\dfrac{(s+\omega+\omega_m)^{1/2}}{(s+\omega_m)^{1/2}+(s+\omega+\omega_m)^{1/2}} \; e^{-\alpha_\omega(s+\omega_m)\left\vert x_{tg}-x_{0}\right\vert
/v_{f}}, & \quad x_{0}\neq x_{tg} \\[4mm]
\dfrac{1}{s} \dfrac{(s+\omega+\omega_m)^{1/2}-(s+\omega_{m})^{1/2}}{(s+\omega+\omega_m)^{1/2}+(s+\omega _{m})^{1/2}}, & \quad x_{0}=x_{tg}.
\end{cases}
\label{twocases}
\end{equation}

Let us first focus on the case where the searcher starts at the target location  ($x_{0}=x_{tg}$). Inverting the equation for the Laplace transform by means of the Faltung theorem and the shift theorem, one obtains

\begin{equation}
1-S^{\ast}(t)=\int_0^t d\tau \, e^{-\omega_m \tau} {\cal L}^{-1}_{s\to \tau}
\left\{\frac{((s+\omega)^{1/2}-s^{1/2})^2}{\omega}\right\}.
\end{equation}

According to ref. \cite{RobertsKaufman} (p. 215, formula 79), one has

\begin{equation}
{\cal L}^{-1}_{s\to \tau}=
\left\{\frac{((s+\omega)^{1/2}-s^{1/2})^2}{\omega}\right\}=\frac{1}{\tau} \,\, e^{-\omega \tau/2} I_1\left(\frac{\omega \tau}{2}\right),
\end{equation}

where $I_1(\cdot)$ stands for a modified Bessel function. We thus find

\begin{equation}
1-S^{\ast}(t)=\int_0^t d\tau \, \tau^{-1} e^{-(\omega+2\omega_m)\tau/2} I_1\left(\frac{\omega \tau}{2}\right).
\label{KPxteqx0}
\end{equation}

One can easily check with look-up tables that the value of the integral on the right hand side in the limit $t\to \infty$ coincides with the value of $1-S^\ast_\infty$ given in Eq. \eqref{effinf}. In passing, we also note that in the absence of mortality ($\omega_m=0$) the integral is known for all times $t$ (see e.g. ref. \cite{Prudnikov2}, p. 46). One finds

\begin{equation}
1- S(t)=1-e^{-\omega t/2} I_1\left(\frac{\omega t}{2}\right)-e^{-\omega t/2}I_0\left(\frac{\omega t}{2}\right).
\end{equation}

Returning to the general case $\omega_m \neq 0$, the killing probability can be rewritten as follows:

\begin{align}
1-S^{\ast}(t)& =\dfrac{\alpha_\omega (\omega_m)-\omega _{m}}{\alpha_\omega (\omega_m)+\omega _{m}}-
\int_t^\infty d\tau \,\tau^{-1} e^{-(\omega+2\omega_m)\tau/2} I_1\left(\frac{\omega \tau}{2}\right).
\end{align}

For long (but finite) times $t$, we now use in the above formula the asymptotic expansion for large $x$:

\begin{equation}
I_1(x)=\frac{e^x}{\sqrt{2\pi x}}\left\{1-\frac{3}{8x}-\frac{15}{128x^2}+O(x^{-3})\right\}.
\label{expI1}
\end{equation}

Neglecting all the terms beyond the first subdominant term and making use of partial integration we find

\begin{align}
\int_t^\infty d\tau \,\tau^{-1} e^{-(\omega+2\omega_m)\tau/2} I_1\left(\frac{\omega \tau}{2}\right)
&\approx \frac{1}{\sqrt{\pi}} \int_t^\infty d\tau \,\tau^{-1}\frac{e^{-\omega_m \tau}}{(\omega \tau)^{1/2}}-
\frac{3}{4\sqrt{\pi}}\int_t^\infty d\tau \,\tau^{-1}\frac{e^{-\omega_m \tau}}{(\omega \tau)^{3/2}} \nonumber \\
& = \frac{1}{\sqrt{\pi\omega}}\frac{e^{-\omega_m t}}{\omega_m  t^{3/2}}-\frac{3}{\sqrt{\pi\omega}}\frac{e^{-\omega_m t}}{(2\omega_m+4\omega)\omega_m t^{5/2}}
+O\left(\frac{e^{-\omega_m t}}{t^{7/2}} \right).
\end{align}
At long times one thus has

\be
1-S^{\ast}(t)=\dfrac{\alpha_\omega (\omega_m)-\omega _{m}}{\alpha_\omega (\omega_m)+\omega _{m}}-\frac{1}{\sqrt{\pi\omega}}\frac{e^{-\omega_m t}}{\omega_m  t^{3/2}}+\frac{3}{\sqrt{\pi\omega}}\frac{e^{-\omega_m t}}{(2\omega_m+4\omega)\omega_m t^{5/2}}
+O\left(\frac{e^{-\omega_m t}}{t^{7/2}} \right).
\ee

In the opposite limit of early times we can make use of the Taylor expansion

\be
I_1(x)=\frac{x}{2}+\frac{x^3}{16}+O(x^5)
\label{earlyexpI1}
\ee
and the series definition of the exponential function in \eqref{KPxteqx0} to obtain

\be
1-S^{\ast}(t)=\frac{\omega}{4}\,\int_0^t d\tau \,[1-\left(\frac{\omega+2\omega_m}{2}\right)\tau+
\left(\frac{\omega^2}{32}+\frac{(\omega+2\omega_m)^2}{8}\right)\tau^2+O(\tau^3)].
\ee
Thus,

\be
1-S^{\ast}(t)=\frac{\omega t}{4}-\frac{\omega(\omega+2\omega_m)}{16}\,t^2+
\left(\frac{\omega^3}{384}+\frac{\omega(\omega+2\omega_m)^2}{96}\right)t^3+O(t^4).
\ee

We now turn to the case $x_0 \neq x_{tg}$. Taking the inverse Laplace transform of the expression given in \eqref{twocases} for this case, one finds

\begin{equation}
1-S^{\ast}(t)=\omega^{-1} \int_0^t d\tau \, e^{-\omega_m \tau} {\cal L}^{-1}_{s\to \tau}
\left\{(s+\omega-\alpha_\omega(s))e^{-\alpha_\omega(s)\tau_0}\right\},
\label{KParbt}
\end{equation}
where we have introduced $\tau_0=v_f^{-1}|x_{tg}-x_0|$. In what follows we restrict ourselves to the case $t \ge \tau_0$, since for $t<\tau_0$ one trivially has $1-S^{\ast}(t)\equiv 0$. Now we invoke the following results given in ref. \cite{FoongKano94}:

\begin{align}
{\cal L}^{-1}_{s\to \tau}\left\{e^{-\alpha_\omega(s)\tau_0}\right\}&=e^{-\omega \tau/2}\delta(\tau-\tau_0)+\frac{\omega \tau_0}{2}\frac{e^{-\omega \tau/2}}{\sqrt{\tau^2-\tau_0^2}}\,I_1\left(\frac{\omega}{2}
\sqrt{\tau^2-\tau_0^2}\right)\theta(\tau-\tau_0), \\
{\cal L}^{-1}_{s\to \tau}\left\{s\,e^{-\alpha_\omega(s)\tau_0}\right\}&
=\frac{\partial}{\partial \tau}\left\{e^{-\omega \tau/2}\delta(\tau-\tau_0)+
\frac{\omega \tau_0}{2}\frac{e^{-\omega \tau/2}}{\sqrt{\tau^2-\tau_0^2}}\,I_1\left(\frac{\omega}{2}
\sqrt{\tau^2-\tau_0^2}\right)\theta(\tau-\tau_0)\right\}, \\
{\cal L}^{-1}_{s\to \tau}\left\{\alpha_\omega(s)\,e^{-\alpha_\omega(s)\tau_0}\right\}&=
e^{-\omega \tau/2}\left(\frac{\partial^2}{\partial \tau^2}-\frac{\omega^2}{4}\right)\,I_0\left(\frac{\omega}{2}
\sqrt{\tau^2-\tau_0^2}\right)\theta(\tau-\tau_0),
\end{align}
where $\theta(\cdot)$ stands for the Heaviside step function. Substituting these results into Eq.~\eqref{KParbt}, applying partial integration for integrals involving derivatives of Bessel functions and carrying out integrals not involving Bessel functions,
we are left with the following expression:

\begin{align}
1-S^{\ast}(t) =&\left(1+\frac{\omega_m}{\omega}\right)\,e^{-(\omega+2\omega_m)\tau_0/2}
-\left(\frac{(\omega+2\omega_m)^2}{4\omega}-\frac{\omega}{4}\right)
\int_{\tau_0}^t d\tau \, \,e^{-(\omega+2\omega_m)\tau/2} I_0\left(\frac{\omega}{2}\sqrt{\tau^2-\tau_0^2}\right) \nonumber \\
&+\left(\frac{\omega+\omega_m}{2}\right)\int_{\tau_0}^t d\tau \, \frac{\tau_0}{\sqrt{\tau^2-\tau_0^2}} \,e^{-(\omega+2\omega_m)\tau/2} I_1\left(\frac{\omega}{2}\sqrt{\tau^2-\tau_0^2}\right)
\nonumber \\
&-\left(\frac{\omega+2\omega_m}{2\omega}\right)e^{-(\omega+2\omega_m)t/2} I_0\left(\frac{\omega}{2}\sqrt{t^2-\tau_0^2}\right)
-\frac{1}{2}\sqrt{\frac{t-\tau_0}{t+\tau_0}}\,e^{-(\omega+2\omega_m)t/2}
I_1\left(\frac{\omega}{2}\sqrt{t^2-\tau_0^2}\right)
\label{kpxtdiffx0}
\end{align}
where additional use of the identity $(\partial/\partial x) I_0(x)=I_1(x)$ has been made to express the time derivative of the zeroth-order modified Bessel function in terms of the first order modified Bessel function.

Note that there is a discontinuity of $1-S^{\ast}(t)$ at $t=\tau_0$, where it jumps from zero to a finite value. In this case, $1-S^{\ast}(\tau_0)$ is the probability that the target is found by the creeper exactly at $t=\tau_0$. For this to happen, \emph{i}) the creeper must start moving towards the target at time $t_0$, \emph{ii}) it must not change direction in the time interval between $0$ and $\tau_0$ and, \emph{iii}) it must not not die during this time interval. The probabilities for these three independent events are respectively $1/2$, $e^{-\omega \tau_0/2}$ and $e^{-\omega_m \tau_0}$. The product of the three yields the result $1-S^{\ast}(\tau_0)=(1/2) \, e^{-(\omega+2\omega_m)\tau_0/2}$.

Let us now focus on the study of the asymptotic behavior of \eqref{kpxtdiffx0}. We first start with the long time regime. Making use of the $x$-large expansion

\begin{equation}
I_0(x)=
\frac{e^x}{\sqrt{2\pi x}}\left\{1+\frac{1}{8x}+\frac{9}{128x^2}+O(x^{-3})\right\}
\label{expI0}
\end{equation}
together with \eqref{expI1}, and applying partial integration to the corresponding integrals, one finds the long time behavior:

\begin{align}
1-S^\ast(t)=&\frac{\alpha_\omega (\omega_m) }{\alpha_\omega (\omega_m)+\omega _{m}} \, e^{-\alpha_\omega(\omega_m)\tau_0}-\left(\frac{1+(\omega+\omega_m)\tau_0}{2\omega_m\sqrt{\pi \omega}}\right)\,\frac{e^{-\omega_m t}}{t^{3/2}} \nonumber \\
& +\left(\frac{3(2\omega+\omega_m)(1+(\omega+\omega_m)\tau_0)}{8\sqrt{\pi \omega^3}\omega_m^2}\right)\,\frac{e^{-\omega_m t}}{t^{5/2}}+O\left(\frac{e^{-\omega_m t}}{t^{7/2}}\right).
\end{align}

Let us now consider the opposite limit of short times. Here, one uses

\be
I_0(x)=1+\frac{x^2}{4}+\frac{x^4}{64}+O(x^6)
\label{earlyexpI0}
\ee
together with \eqref{earlyexpI1} to perform an expansion of the functions appearing in \eqref{kpxtdiffx0} in the small parameter $\Delta=t-\tau_0$. The final result is

\be
1-S^\ast(t)= e^{-(\omega+2\omega_m)\tau_0/2}\left(\frac{1}{2}+\frac{\omega(\omega\tau_0+2)}{16}\Delta
+\frac{(\omega\tau_0-4)\omega^3\tau_0-16\omega^2-16(\omega\tau_0+2)\omega\omega_m}{512}\,\Delta^2+
O(\Delta^3)\right).
\ee

Note that the linear term in $\Delta$ does not depend on $\omega_m$. The mortality rate only appears in the quadratic order term.

Finally, we note that in the absence of mortality, the formula \eqref{kpxtdiffx0} takes the simplified form

\begin{align}
1-S(t) =&e^{-\omega\tau_0/2}-\frac{1}{2}\,e^{-\omega t/2} I_0\left(\frac{\omega}{2}\sqrt{t^2-\tau_0^2}\right)
-\frac{1}{2}\sqrt{\frac{t-\tau_0}{t+\tau_0}}\,e^{-\omega t/2}
I_1\left(\frac{\omega}{2}\sqrt{t^2-\tau_0^2}\right)
 \nonumber \\
&+\frac{\omega}{2}\int_{\tau_0}^t d\tau \, \frac{\tau_0}{\sqrt{\tau^2-\tau_0^2}} \,e^{-\omega \tau/2} I_1\left(\frac{\omega}{2}\sqrt{\tau^2-\tau_0^2}\right).
\end{align}
Alternatively, this formula can be obtained as the spatial integral of the propagator given in ref. \cite{MasPorrWeiss} for the telegrapher's equation in the presence of a single trap.

\begin{figure}
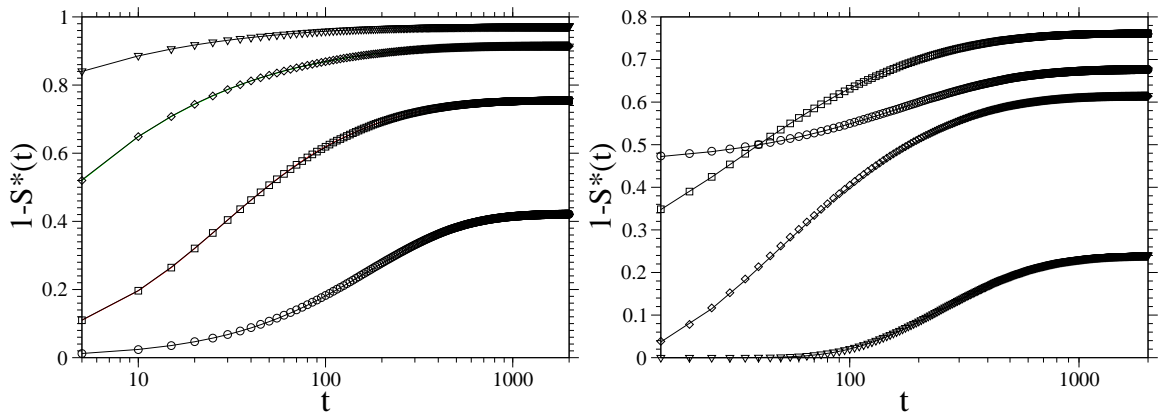

\includegraphics[scale=0.3]{xteqx0logmod.eps}
\includegraphics[scale=0.3]{xtdiffx0logmod.eps}
\caption{Log-linear plot of the cumulative probability to find the target as a function of time. The left figure corresponds to the case $x_{tg}=x_0$, whereas the right figure corresponds to the case $|x_{tg}-x_0|=10$. In both cases we take $\omega_m=0.002$. The symbols correspond to simulation results for $\omega=0.01$ (circles), $\omega=0.1$ (squares), $\omega=1$ (diamonds) and $\omega=10$ (triangles). The solid lines correspond to the analytic results respectively given by the formulae \eqref{KPxteqx0} and \eqref{kpxtdiffx0}.}
\label{figtimedep}
\end{figure}

Returning to the case with mortality, Fig. \ref{figtimedep} shows a comparison for different parameter values between analytical and numerical results for the detection probability as a function of time. Excellent agreement is found.

We close by noting that, in the presence of creeper mortality, the survival probability of the target $S^\ast(t)$ and that of the creeper $S_{cr}^\ast(t)$ become different if one assumes that both creeper and target die instantaneously upon detection. However, there is a straightforward relation between both of them, namely,   $S^\ast (t)=S^\ast_{cr}(t)+\omega_m \int_0^t S^\ast_{cr}(\tau) d\tau$ (see ref. \cite{SurvivalEvanescent} for a detailed derivation). Obviously, one must have 
$S^\ast(t)>S_{cr}^\ast(t)$ in this case, since the creeper can die not only upon encounter with the target, but also spontaneously. 

\subsubsection{$N$ mortal creepers}

Consider the case of $N>1$ independent creepers that begin their walk at the same location and at the same time.
The question in this case is whether an optimal search efficiency of the walkers (i.e., minimum value of the survival probability of the target) is still found. The search efficiency is now simply $1-S_{N,\infty} =1-(S_\infty^\ast)^N$. The condition for $\omega_\text{opt}$ is again found by requiring that the derivative of the efficiency with respect to $\omega$ be zero. However, the condition for this derivative to become zero is the same as for the $N=1$ case. Therefore, the value of $\omega_\text{opt}$ is the same. However, since, $S_\infty^\ast<1$, the detection probability of the target tends to one as $N\to \infty$. In words, when the number of walkers becomes infinitely large, one of them will sooner or later reach the target with certainty no matter how quickly the walkers die on average.

\subsection{A finite domain}
 \label{subs:findom}

\subsubsection{Single mortal creeper}

The most straightforward extension of our results to a finite system is to a ring of length $L$. On this ring, we place a point target and calculate its survival probability  $S_{\infty }^{\ast }$ in the presence of a single mortal creeper. This is most easily done by exploiting the relation between the
probability density $p(x,t|x_0)$ introduced in Eq.~(\ref{pjeq}) and that of the finite system $p_L(x,t|x_{0})$:
\begin{equation}
p_L(x,t|x_{0})=\sum_{n=-\infty}^\infty p(x+nL,t|x_{0}), \qquad 0<x<L.
\end{equation}
In Laplace space, this translates into the relation
\begin{equation}
p_L(x,s|x_0)=\sum_{n=-\infty}^\infty p(x+nL,s|x_0), \qquad 0 \le x \le L.
\end{equation}
Using Eq.~(\ref{LTp2}) we then obtain an expression in terms of a geometric series which can be summed.  The sum yields
\begin{equation}
p_L(x,s|x_0)=\frac{1}{2v_f}\frac{\alpha_w(s)}{s} \,\left\{\frac{e^{-\alpha_w(s)\,|x-x_0| /v_f}+e^{-\alpha_w(s)\,(L-|x-x_0|)/v_f}}{1-e^{-\alpha_w(s)\,L/v_f}}\right\}, \qquad 0 \le x \le L.
\label{PropRing}
\end{equation}
One can easily check that the above equation is normalized, i.e., $\int_0^L dx \, p_L(x,s|0)=s^{-1}$. The limit of an infinite system is also recovered as one lets $L\to \infty$ in the above expression.

The corresponding detection rate $q_L(x_{tg},t|x_0)$ is obtained by making use of Eq. (\ref{PropRing}) in Eq.  (\ref{qrate}). The survival probability of the target is subsequently computed from Eq. (\ref{mortalSP}), which remains valid for a system of finite size. The final result reads as follows:
\begin{equation}
S_{\infty }^{\ast }=
\begin{cases}
1-\dfrac{ \alpha_{\omega}(\omega_{m})  }{\omega _{m} \beta_L^- +\alpha_{\omega}(\omega_{m}) \beta_L^+ }\left(  e^{-\alpha_{\omega}(\omega_{m}) \vert x_{tg}-x_{0} \vert /v_{f}} + e^{-\alpha_{\omega}(\omega_{m}) (L- \vert x_{tg}-x_{0} \vert) /v_{f}}\right),
& \quad x_{0}\neq x_{tg} \\[4mm]
\dfrac{2\omega _{m}\beta_L^- }{\omega _{m} \beta_L^-+\alpha_{\omega}(\omega_{m}) \beta_L^+ }, & \quad x_{0}=x_{tg},
\end{cases}
\label{survivalfinite}
\end{equation}
with $\beta_L^\pm=1\pm e^{-\alpha_{\omega}(\omega_{m}) L/v_{f}}$.

\begin{figure}
\includegraphics{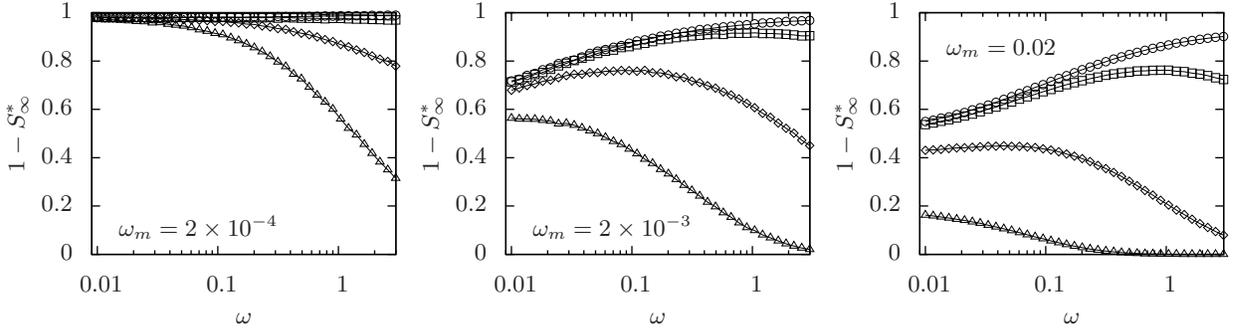}
\caption{Plot of the search efficiency in a finite domain ($L=200$) as a function of $\omega$ for different values of the mortality rate and for different initial conditions: $\vert x_{tg}-x_{0} \vert=0.1$ (circles), $1$ (squares), $10$ (diamonds), and $50$ (triangles). Solid lines correspond to the analytic result (\ref{survivalfinite}) and symbols are obtained numerically from the average over the trajectories of $10^{5}$ independent random creepers. In all cases we take $v_{f}=1$.}
\label{figura2}
\end{figure}

We note the great qualitative and even quantitative similarity between the results for $L$ finite (Figure~\ref{figura2}) and those in Figure~\ref{figura1}. This similarity is also seen quite explicitly when comparing Eqs.~(\ref{survivalfinite})  and (\ref{survivalinf}) in the limits $\omega_m\to 0$ and $\omega_m\to\infty$. The differences between the figures and equations of course arise from our domain size ($L=200$) which is small enough to produce non-negligible finite size effects in the ranges of values we chose for the other parameters.  Without loss of generality we place the target at $x_{tg}=0$ or, equivalently, $x_{tg}=L$.  A physical sense of the situation is obtained as follows. Suppose that the creeper begins its walk at a location near the target (the greatest initial distance between creeper and target occurs when the creeper begins its motion near $x_0=L/2$).
If the creeper moves toward the target along the shorter direction without a direction reversal, it reaches the target very quickly and the target does not survive these events, thus decreasing the survival probability $S_\infty^*$.  These events are independent of system size, and thus are not affected by the finite size of our system.  If the creeper moves toward the target along the long side, the minimum time needed for the creeper to reach the target at $x_{tg}=L$ is of order $L/v_f$.  The mean time to death of the creeper is of order $1/\omega_m$.  If this time is shorter than $L/v_f$, the creeper dies before reaching the target, and finite system size effects are again not seen on average.  This description fits that shown in the right panel of Figure~\ref{figura2}, since $L/v_f = 200$ for the parameter choices indicated in the caption, and $1/\omega_m = 50$.  If the creeper suffers many trajectory reversals, then it is even more likely that it dies before reaching the target.  The left panel is the one in which the parameter choices would most clearly lead to finite size effects since now $1/\omega_m=5000$ and for sufficiently low reversal frequency the searcher will almost certainly reach the target, even though its initial position is not so close to the target.  The survival probability of the target $S_\infty^*$ should be smaller than in the right panel, and this is what we see.  However, the results for the infinite system in Figure~\ref{figura1} and for the finite system in Figure~\ref{figura2} are rather similar, not only qualitatively but even quantitatively for all the parameters considered.

Earlier we introduced Figure~\ref{figura3}, which shows the reversal frequency $\omega$ of the creeper trajectory at which the search efficiency $1-S_\infty^*$ is a maximum as a function of the mortality rate of the creeper.  We called this frequency the optimal persistence $\omega=\omega_\text{opt}$. We see there that the optimal persistence depends quite sensitively on $L$ in some regimes.  For very small systems the optimal value is always near zero (as happens for $L=100$ in Figure~\ref{figura3}), that is, regardless of the mortality rate of the walker the best strategy is ballistic motion. As $L$ increases, a new regime emerges where $\omega_\text{opt}$ is non-zero, that is, the creeper is most likely to step on the target with reversals in the direction of motion.  For very high creeper mortality rate, the best strategy is again ballistic motion. This is consistent with the behavior seen in Figure~\ref{figura2}. Note that the maximum value of $\omega_\text{opt}$ is roughly the same for all the $L$'s shown in Figure~\ref{figura3} but the width of $\omega_m$-interval where $\omega_\text{opt}$ is appreciably different from zero is strongly $L$-dependent and it is narrower for smaller $L$. Finally, as $L\rightarrow \infty$ the region with a finite optimal persistence increases in the direction of low mortality and eventually becomes dominant.

\begin{figure}[t]
\includegraphics[scale=0.6,angle=270]{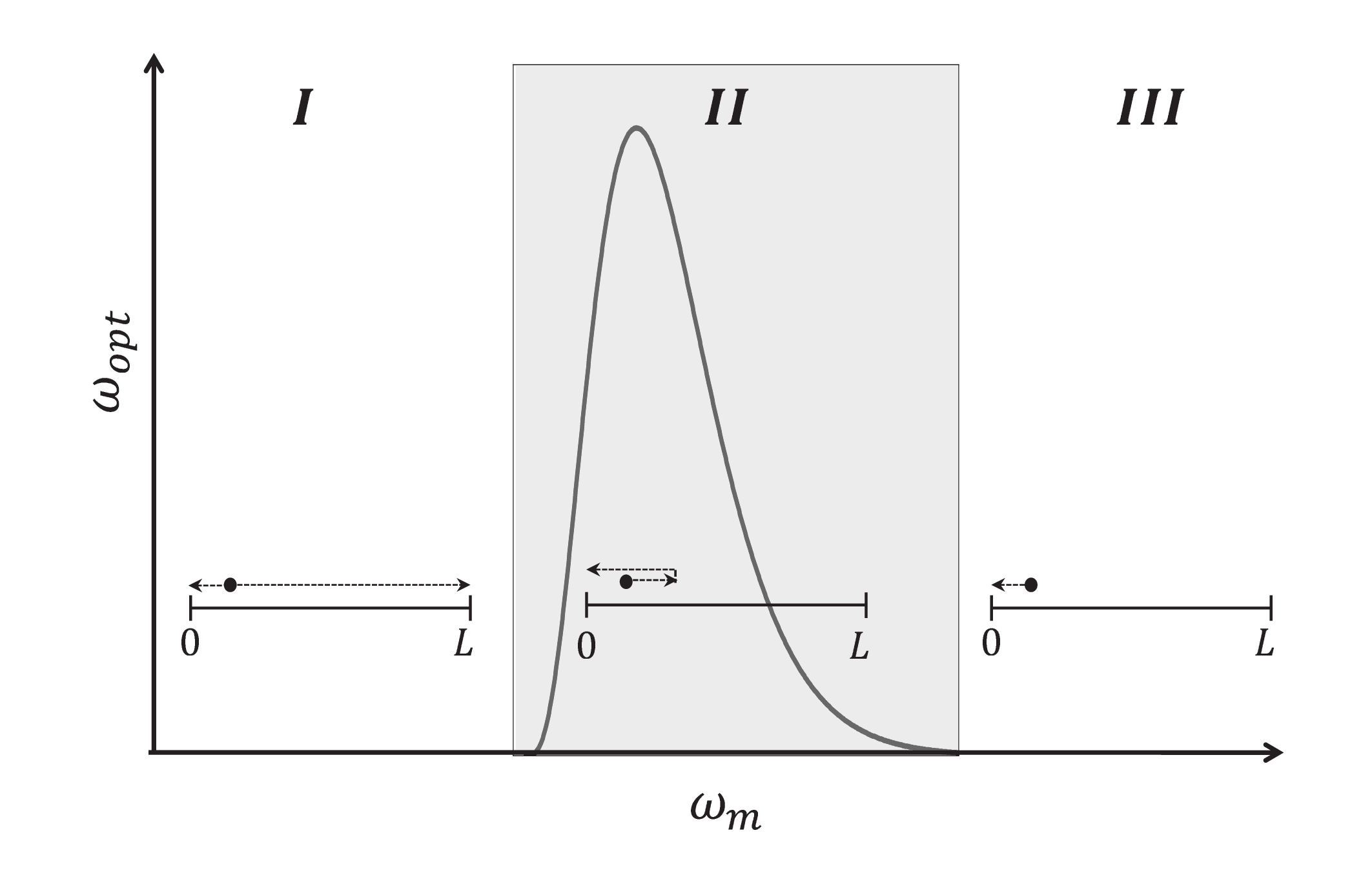}
\caption{Schematic representation of the three regimes observed in Figure~\ref{figura3} for intermediate values of $L$. Sketches of typical trajectories
showing how the creeper (the black point) most likely reaches the target are shown as arrows for each region. The target is located at $x_{tg}=0$, which is equivalent to $x_{tg}=L$ because of the periodic boundary conditions.}
\label{figura4}
\end{figure}

In order to further clarify these results, Figure \ref{figura4} shows a sketch of typical trajectories of the creeper in the three regimes observed for intermediate values of $L$. For a high mortality rate (Region III), the only way the searcher can successfully hit the target is to move ballistically toward it along the shortest path before dying, and therefore maximum persistence is desirable. However, as the mortality rate decreases, the creeper has time to reverse its trajectory and move in one direction or the other, eventually with equal probability. Introducing some breaking of persistence in this regime is seen to be a desirable strategy for the creeper. The reason is that in those runs where the creeper starts moving away from the target, a change in direction makes it more likely to detect the target before dying. Finally, if mortality is so low (Region I) that the searcher has enough time to reach the target location $x=L$ by moving ballistically before dying, one again finds that $\omega_{opt}=0$. Breaking the persistence in this regime would evidently introduce unnecessary overlapping of the trajectory with itself and increase the cumulative probability of death due to the additional delay. As one would expect, in the limit $L \rightarrow \infty$ Region II grows at the expense of region I (which becomes vanishingly small), and $\omega_{opt}$ saturates at a value that depends on the initial separation $|x_{0}|$ (recall that we have set $x_{tg}=0$ for simplicity).

From our analysis so far, it is clear that the transition from Region II to Region III
points to the fact that the characteristic lifetime $\omega_{m}^{-1}$ of the creeper has gone from being longer to being shorter than $x_{0}/v_{f}$, while the transition from Region II  to Region I occurs as $\omega_{m}^{-1}$ grows and becomes comparable to $L/v_{f}$, the typical time it takes the creeper to reach the target if the motion is ballistic. Implicit in this reasoning  is the condition $x_{0} \ll L/2$; otherwise, the necessary trade-off for the onset of an optimum persistence breaks down. This is the reason why in Figure \ref{figura3} no onset of Region II is observed for sufficiently small values of $L$.

Mathematically, the specific condition that ensures the existence of Region II so that the search efficiency is not always maximal for $\omega=0$ arises from the requirement $d (1-S_{\infty^*})/ d \omega \vert _{\omega=0} >0$. Using Eq.~(\ref{survivalfinite}) in this requirement one finds that the condition
\begin{equation}
\frac{2\omega_{m}}{v_{f}} \left( L- x_{0} +x_{0} e^{\omega_{m} (L-2x_{0}/v_{f})} \right) +e^{-2\omega_{m}x_{0}/v_{f}} + e^{-\omega_{m}L/v_{f}} - e^{\omega_{m} (L-2x_{0})/v_{f}}-1 <0
\end{equation}
must be fulfilled.  From this condition one can establish that both $\omega_{m} \gg L/v_{f}$ and $x_{0} \ll L/2$  must necessarily hold for Region II to exist.  Summarizing, a non-zero optimal persistence can only arise if the searcher is initially within a short distance of the target. Distant targets can not be reached within the creeper's lifetime.

\subsubsection{$N$ mortal creepers and the target problem}

Next we again consider a target in the presence of $N$ creepers.  If the creepers start from the same location, the remarks made in the case of an infinite system apply here as well. More interesting is the situation where the position of each creeper is drawn from a uniform distribution
$p(x_0)=L^{-1}$ in the interval $[-L/2,L/2]$, with the target located at the center of the interval ($x_{tg}=0$). This so-called target problem has long been discussed in the literature for immortal random walkers and more recently by three of us for mortal diffusing particles and mortal walkers with independent stepping time and stepping length distributions \cite{SurvivalEvanescent, AYLJMMNP, MortalBookChapter, NovaBC}. The survival probability of the target at time $t$ in the presence of the mortal creepers is given by the same expression as for usual walkers (leapers), namely,
\begin{equation}
S_{N}^\ast(t)=\left(\frac{1}{L} \int_{-L/2}^{L/2} dx_0 \, S^\ast(t|x_0) \right)^N=\left(\frac{2}{L} \int_{0}^{L/2} dx_0 \, S^\ast(t|x_0) \right)^N.
\end{equation}
In the thermodynamic limit we let both $L$ and $N$ go to infinity, keeping the initial density of mortal creepers fixed at a value
$\rho_0=N/L$. The above equation then becomes
\begin{equation}
\bar{S}^\ast(t) \equiv \lim_{N\to\infty} S_{N}^\ast(t)=\lim_{N\to\infty} \left[1-\frac{2\rho_0}{N}\int_0^{N/(2\rho_0)} dx_0 \, (1-S^\ast(t)) \right]^N=
\mbox{exp}\{-2\rho_0 \int_0^\infty dx_0\, (1-S^\ast(t))\},
\label{Sastt}
\end{equation}
where $\bar{S}^\ast(t)$ denotes the survival probability of the target in the aforementioned thermodynamic limit.
For $t\to\infty$, we now use Eq. \eqref{effinf} to evaluate the last integral explicitly:
\begin{equation}
\bar{S}_\infty^\ast =\lim_{t\to\infty} \bar{S}^\ast(t) = \mbox{exp}\{-2\rho_0 v_f/(\omega_m+\alpha_\omega(\omega_m))\}.
\label{lastterm}
\end{equation}
Note that the corresponding search efficiency $1-\bar{S}_\infty^\ast$  decreases steadily with increasing average direction reversal frequency $\omega$, implying that no optimal persistence $\omega_\text{opt}>0$ arises in this limit.

\begin{figure}
\includegraphics[scale=0.3]{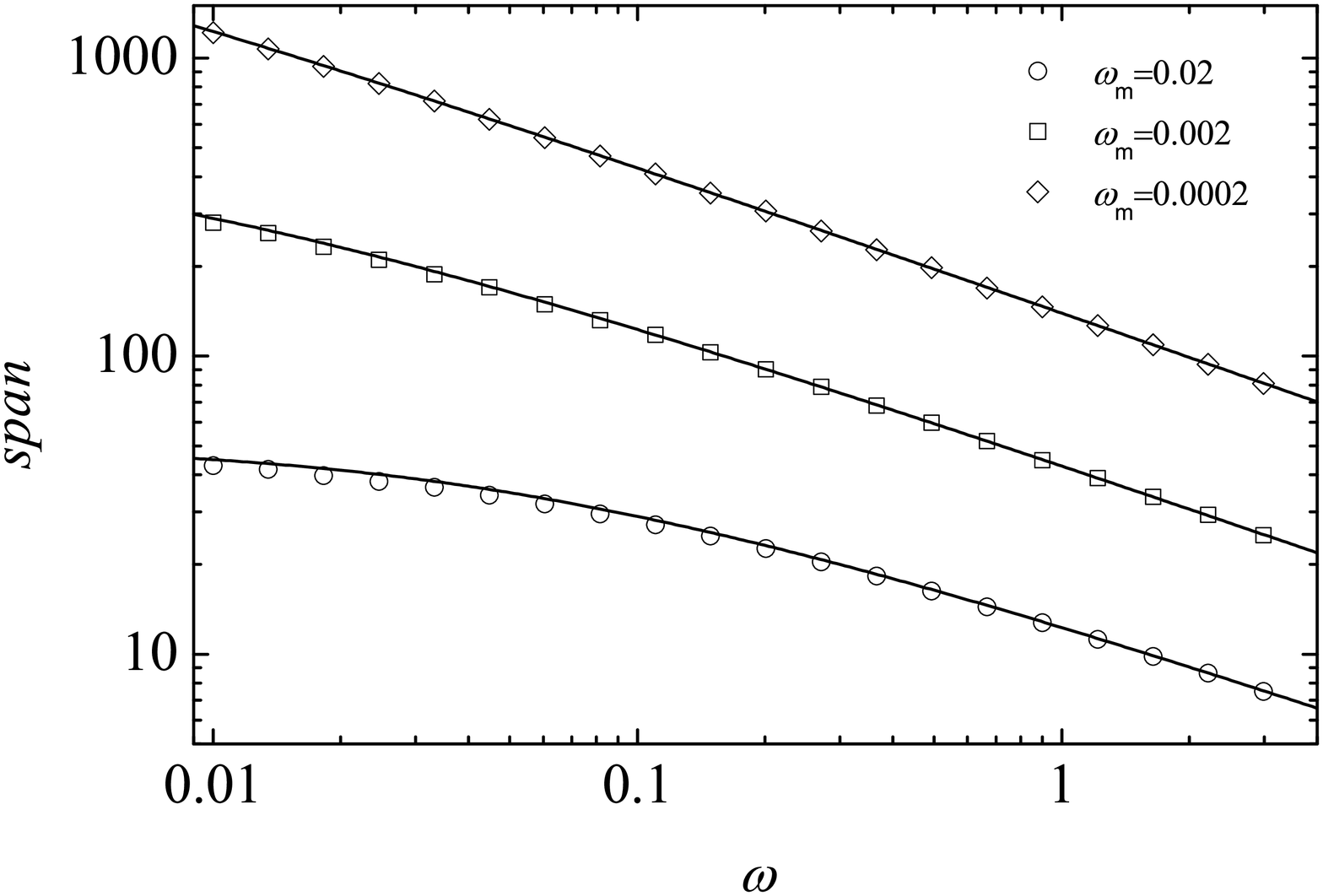}
\caption{Asymptotic span of a $1d$ mortal creeper as a function of the frequency of reorientation for different mortality rates (we have set $v_f=1$). }
\label{figspan}
\end{figure}

There is an interesting and useful connection between the last term of Eq.~(\ref{Sastt}) and the territory explored by a single walker during time $t$
\cite{ BluKlaZuOptical, ourPRL}.  Explored territory includes all those locations visited any number of times.  For mortal creepers, this territory is given by the integral in the argument of the exponential in Eq.~(\ref{Sastt}), that is, $2\int_0^\infty dx_0\, (1-S^\ast(t))$. Thus, according to Eq.~\eqref{lastterm}, in the limit $t\to \infty$ the territory explored by a single mortal creeper is $2v_f/(\omega_m+\alpha_\omega(\omega_m))$ (in $1d$ this quantity is identical with the span of the mortal random walk in the absence of the target). This analytic result is confirmed by numerical simulations (see Fig. \ref{figspan}).

\section{Results in two dimensions}
\label{sec4}

In order to assess whether the onset of an optimal persistence is specific to the one-dimensional case and the peculiarities of space exploration properties of random creepers in this constrained geometry, we have briefly investigated the behavior of the search efficiency in a two-dimensional system. The route to an analytic solution appears to be cumbersome, but numerical simulations allow us to conclude that an optimal persistence also exists in this case.

We again assume perfect detection, implying that the target will instantaneously be detected whenever the trajectory of a pointwise searcher attempts to penetrate it. Turns interrupting the ballistic trajectories of the searchers occur according to a uniform turn angle distribution. In particular, this also applies for the velocity initial condition.

\begin{figure}
\includegraphics{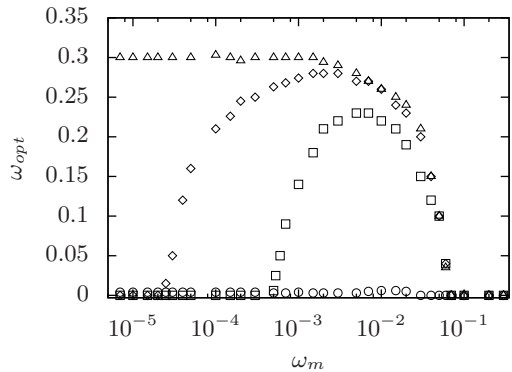}
\caption{Optimal persistence obtained as a function of the mortality rate $\omega_{m}$ for periodic square lattices of size $L \times L$, with $L=25$ (circles), $100$ (squares), $400$ (diamonds) and $L \rightarrow \infty$ (triangles). The results were obtained by averaging over trajectories of $10^{5}$ independent mortal creepers. The target is assumed to be a circle with unit radius centered at the origin ($\vec{x}_t=(x_{tg},y_t)=0$). In all cases the parameter values $\sqrt{(x_{tg}-x_{0})^{2}+(y_{t}-y_{0})^{2}}=2$ and $v_{f}=1$ were taken.}
\label{figura5}
\end{figure}

Results for the search efficiency in two-dimensional torii (arising from periodic boundary conditions) are displayed in Figure \ref{figura5}. Trends similar to those of the one-dimensional case displayed in Figure~\ref{figura3} are observed. Once again, regions I, II and III can be clearly identified and the behavior of the optimal persistence as a function of $L$ follows the same trends as in dimension $d=1$. A reasoning similar to the one underlying Figures~\ref{figura3} and \ref{figura4} is expected to apply here as well.

\section{Summary and Outlook}
\label{sec5}

We have investigated the effect of mortality processes in the efficiency of a search process performed by a class of randomly moving particles with spatiotemporal coupling. Our main conclusion is that for a suitable parameter choice a finite optimal persistence maximizing the long-time detection probability of the target appears. This optimization effect has been shown to exist both in finite and infinite $1d$ domains, and to persist in the $2d$ case (in the case of an infinite one-dimensional domain we have also derived an explicit expression for the survival probability of the target after a finite time $t$). However, optimization with respect to the frequency of turns is suppressed in the limit where the target is surrounded by a finite density of mortal creepers initially distributed at random over an infinite interval ($1d$ target problem).

Even though our model is a strong simplification of real systems (say, a predator hunting a prey), the robustness of the aforementioned effect leads us to believe that it may also play a major role in more realistic systems. In any case, our model incorporates features which are likely to be relevant in an ecological context, namely, the fact that long jumps must pay a time penalty, and lifetimes which are comparable to the relevant time scales of transport \cite{campos14b}.

The present work can be extended in many interesting ways. Further efforts are indeed needed to completely ascertain the role of spatiotemporal coupling and mortality in a broad class of target search problems. As already mentioned in the introduction, one could for instance replace the exponentially decreasing waiting time density with a long-tailed waiting time density and investigate whether the resulting walk leads to similar optimization effects as the ones observed here. Such problems are interesting even in the absence of mortality. Bearing in mind possible applications in ecology, the case of L\'evy flights with long-tailed jump length distributions but no coupling between jump lengths and waiting times could be investigated (in this case special care should be taken when defining the details of the detection process). Finally, one could relax the condition of an immobile target and assess the behavior of the search efficiency by devising suitable approximations for the case where the target is also allowed to move. The question is whether some kind of Pascal principle applies, as in the case of conventional Brownian motion \cite{Moreau}.

\section{Acknowledgements}

This work was partially funded by MINECO (Spain) through Grants No. FIS2013-42840-P (partially financed by FEDER funds) (E. A. and S. B. Y.), and FIS-2012-32334 (D. C. and V. M.). Additional financial support was provided by the Generalitat de Catalunya through Grant No. SGR 2014-923 (D. C. and V. M.) and by the Junta de Extremadura through Grant No. GR15104 (E. A. and S. B. Y.).   K. L. gratefully acknowledges support of the U.S. Office of Naval Research (ONR) under Grant No. N00014-13-1-0205.

\end{document}